\documentclass[useAMS,usenatbib]{mn2e}
\usepackage{graphicx}
\usepackage{amssymb}
\usepackage[below]{placeins}
\usepackage{natbib}
\usepackage{multirow}
\usepackage{array}
\usepackage{savesym}
\usepackage{amsmath}
\savesymbol{iint}
\savesymbol{iiint}
\savesymbol{iiiint}
\savesymbol{idotsint}
\usepackage{txfonts}
\restoresymbol{TXF}{iint}
\restoresymbol{TXF}{iiint}
\restoresymbol{TXF}{iiiint}
\restoresymbol{TXF}{idotsint}
\bibpunct{(}{)}{;}{a}{}{,}
\def \arcmin{$^{\prime}$}

\def \xmm{{\emph{XMM-Newton}}}
\def \rosat{{\emph{ROSAT}}}
\def \chandra{{\emph{Chandra}}}
\def \asca{{\emph{ASCA}}}

\def \suzaku{{\emph{Suzaku}}}

\newcommand{\ion}[2]{#1\,{\sc{#2}}}

\bibpunct{(}{)}{;}{a}{}{,}

\title[Virgo Cluster out to the virial radius]
{X-ray Spectroscopy of the Virgo Cluster out to the Virial Radius}

\author[O. Urban et al.]{O. Urban$^{1,2}$, N. Werner$^{1}$\thanks{E-mail: norbertw@stanford.edu}\thanks{Chandra/Einstein fellow}, A. Simionescu$^{1}$\thanks{Einstein fellow}, S. W. Allen$^{1}$, H. B\"ohringer$^{3}$\\
$^1$Kavli Institute for Particle Astrophysics and Cosmology, Stanford University, 452 Lomita Mall, Stanford, CA 94305-4085, USA \\
and SLAC National Accelerator Laboratory, 2575 Sand Hill Road, Menlo Park, CA 94025, USA \\ 
$^{2}$\'UTFA, P\v{r}\'irodov\v{e}deck\'a fakulta, Masarykova Univerzita, Kotl\'a\v{r}sk\'a 2, Brno, Czech Republic\\
$^{3}$Max-Planck-Institut f\"ur extraterrestrische Physik, Giessenbachstr. 1, D-85748 Garching, Germany \\
}
\begin{document}
\maketitle
\begin{abstract}
We present results from the analysis of a mosaic of thirteen \xmm\ pointings covering the Virgo Cluster from its center northwards out to a radius $r\sim$1.2~Mpc ($\sim$4.5~degrees), reaching the virial radius and beyond. This is the first time that the properties of a modestly sized ($M_{\mathrm{vir}}\sim1.4\times10^{14}$~M$_{\odot}$, $kT\sim2.3$~keV), dynamically young cluster have been studied out to the virial radius. The density profile of the cluster can be described by a surprisingly shallow power-law $n_{\mathrm{e}}\propto r^{-\beta}$ with index $\beta=1.21\pm0.12$. In the radial range of $0.3r_{\mathrm{vir}}<r<r_{\mathrm{vir}}$, the best fit temperature drops by roughly 60 per cent. Within a radius $r<450$~kpc, the entropy profile has an approximate power-law form $K\propto r^{1.1}$, as expected for gravitationally collapsed gas in hydrostatic equilibrium. Beyond $r\sim450$~kpc, however, the temperature and metallicity drop abruptly, and the entropy profile becomes flatter, staying consistently below the expected value by a factor of 2--2.5. The most likely explanation for the unusually shallow density profile and the flattening of entropy at large radius is clumping in the ICM. Our data provide direct observational evidence that the ICM is enriched by metals all the way to $r_{200}$ to at least $Z=0.1$ Solar. 
\end{abstract}


\section{Introduction}

Sitting at the top of the mass spectrum of the virialized objects in the Universe, galaxy clusters serve as unique probes of cosmological parameters \citep[e. g.][]{schuecker2003,allen2008,vikhlinin2009,mantz2010} and provide an approximately fair sample of the matter content of the Universe \citep{1993Natur.366..429W,1996ApJ...462...32C}. 

Over the past 20 years, progress in understanding the astrophysics of galaxy clusters and their use as probes for cosmology was driven by X-ray observations. However, those observations only targeted the inner part of their volumes extending at most out to $r_{500}$\footnote{Here $r_{\Delta}$ means radius within which the mean enclosed matter density is $\Delta$-times the critical density at redshift $z$.}, where the X-ray emission is brightest; extrapolations were used to infer their properties further out. Until very recently, the thermodynamic properties of the cluster outskirts were not directly observable. 

Early work on the outskirts of clusters focused on determining the surface brightness profiles, using \rosat\ data in particular \citep{1999ApJ...525...47V,2005A&A...439..465N}. This provided density estimates, but essentially no temperature information. 
Reliable information regarding the temperature of the intra-cluster medium (ICM) at the virial radius (which approximately corresponds to $r_{200}$) has become available only recently. Properties of the ICM in the outskirts of a handful of clusters, including PKS~0745-191 \citep{2009MNRAS.395..657G}, Abell~2204 \citep{2009A&A...501..899R}, Abell 1795 \citep{2009PASJ...61.1117B}, Abell~1413 \citep{2010arXiv1001.5133H}, and Abell~1689 \citep{kawaharada2010}, could be studied thanks to the low instrumental background of the Suzaku satellite. 
The most detailed radial profiles of thermodynamic quantities measured all the way out to $r_{200}$ were recently obtained for the Perseus Cluster \citep{simionescu2011}. 
Among other results, these observations provide evidence that the ICM at large radii is highly clumped. 

However, all of these clusters studied with \suzaku\ at large radii to date are hot, massive, relatively relaxed systems. In order to have a complete picture of how galaxy clusters evolve within the context of hierarchical structure formation, we also need to understand less massive, less relaxed systems. Such systems are also critical to understand for cosmological studies: a large fraction of the integrated Sunyaev-Zel'dovich (SZ) power spectrum is expected to come from modestly sized clusters and groups, which are much more common than their more massive counterparts.

The nearest, brightest, modestly sized, dynamically young cluster and therefore the ideal object to study for these purposes is the Virgo Cluster. The cluster is at a distance of 16~Mpc and has an average temperature of $\sim$2~keV. The bulk of the X-ray emission from this relatively cool system arises in the soft X-ray band, where the effective area of current X-ray mirrors is large and the signal-to-detector background contrast is optimal. Therefore, using a careful background treatment, we can study the thermodynamic properties and the chemical abundances of the low surface brightness regions of Virgo out to $r_{200}$, even with the \xmm\ satellite. 

 \citet{1994Natur.368..828B} used \rosat\ Position Sensitive Proportional Counter (PSPC) observations from the ROSAT All-Sky Survey to study the structure of the Virgo Cluster. They discovered that the X-ray emission in general traces the galaxy distribution and confirmed 3 sub-clumps, centered on the galaxies M87, M49, and M86 respectively \citep[reported earlier by][]{binggeli1987,1993A&AS...98..275B}. 
\citet{2001ApJ...549..228S} derived an extensive temperature map using hardness ratio values from \asca\ observations covering an area of 19~deg$^2$. 
For a mean temperature of $kT=2.3$~keV, the scaling relations by \citet{arnaud2005} predict $M_{200}=1.4\times10^{14}$~M$_{\odot}$ and $r_{200}=1.08\,\text{Mpc}$, which is a projected radius of $\sim3.9^{\circ}$ in the sky.

In this work, we study the structure of the Virgo Cluster using a mosaic of \xmm\ pointings, which cover it from the center northwards out to $r_{200}$. The pointings are shown in Fig.~\ref{fig:rosat} as circles over-plotted on the adaptively smoothed \rosat\ PSPC image. 
Sect.~\ref{sect:obsanal} discusses the data reduction and extraction of the data products. In Sect.~\ref{sect:results} we present projected surface brightness, temperature and metallicity profiles and determine the deprojected profiles of temperature and density. Based on these values we calculate the pressure and entropy profiles of the ICM. We also discuss in detail various possible sources of systematic errors and their impact on our results. In Sect.~\ref{sect:discus} we discuss the implications of our results and provide a summary and conclusions in Sect.~\ref{sect:concl}.

All the errors are given at $1\sigma$ (68\%) confidence level unless otherwise stated. We assumed the distance of Virgo as that of the central galaxy M87 \citep[16.1~Mpc,][]{2001ApJ...546..681T}. At this distance, 1\arcmin\ corresponds to $\sim4.65\thinspace\text{kpc}$.

\section{Observations and data analysis}
\label{sect:obsanal}

\begin{table*}
\begin{center}
\caption{Summary of the observations. Columns give the observation identifier, starting date of the observation, \xmm\ revolution of the observation, total and cleaned exposure times for MOS1, MOS2 and pn instruments respectively, and the coordinates of the pointings. The pointings in the table are ordered from the south to the north.}
\begin{tabular}{lcccccccccc}
\hline\hline
\multirow{2}{*}{obs. ID}    & date     & \multirow{2}{*}{rev.}  & \multicolumn{3}{c}{Exp. time [ks]} & \multicolumn{3}{c}{Clean time [ks]} & \multirow{2}{*}{RA [$^{\circ}$]} & \multirow{2}{*}{DEC [$^{\circ}$]}\\
                            & DD/MM/YY &                        & MOS1       & MOS2    & pn        & MOS~1     & MOS~2     & pn          &                     &\\
\hline 
0200920101                  & 11/01/05 & 932                    & 95.36       & 95.37    & --$^{(a)}$& 74.87     & 75.65     & --          & 187.71              & 12.36 \\
0106060101                  & 12/07/01 & 291                    & 9.28        & 9.28     & 5         & 8.75      & 9.04      & 5           & 187.71              & 12.73 \\
0106060201                  & 08/07/02 & 472                    & 8.36        & 8.37     & 5         & 4.65      & 4.51      & 1.7         & 187.71              & 13.06 \\
0106060301                  & 04/07/02 & 470                    & 8.89        & 8.89     & 5.52      & 6.87      & 7.3       & 5.1         & 187.71              & 13.39 \\
0106060401                  & 04/02/02 & 470                    & 11.33       & 11.35    & 7.64      & 10.54     & 10.68     & 6.77        & 187.71              & 13.73 \\
0106060501                  & 06/07/02 & 471                    & 17.38       & 17.42    & 13.87     & 14.3      & 14.7      & 11.58       & 187.71              & 14.06 \\
0106060601                  & 08/07/02 & 472                    & 14.37       & 14.37    & 11        & 9.38      & 9.81      & 8.52        & 187.71              & 14.39 \\
0106060701                  & 05/07/02 & 471                    & 13.85       & 13.87    & 10.5      & 4.89      & 6.09      & 0           & 187.71              & 14.73 \\
0106061401$^{(b)}$          & 13/06/02 & 460                    & 8.23        & 8.26     & 4.88      & 7.84      & 7.92      & 4.88        & 187.71              & 15.06 \\
0106060901                  & 10/06/02 & 458                    & 15.03       & 15.03    & 11        & 11.28     & 11.38     & 7.8         & 187.71              & 15.39 \\
0106061001                  & 06/06/02 & 456                    & 12.39       & 12.18    & 8.89      & 8.58      & 8.45      & 5           & 187.71              & 15.73 \\
0106061101                  & 09/06/02 & 458                    & 15.93       & 15.95    & 11.98     & 8.31      & 8.43      & 4.96        & 187.71              & 16.06 \\
0106061201                  & 09/06/02 & 458                    & 17.57       & 17.65    & 13.52     & 14.19     & 14.36     & 10.22       & 187.71              & 16.39 \\
0106061301$^{(c)}$          & 10/06/02 & 458                    & 15.99       & 16.03    & 12        & 14.7      & 14.39     & 11.33       & 187.71              & 17.89 \\
\hline
\multicolumn{8}{l}{$^{(a)}$ data from pn were not used for the central pointing} \\
\multicolumn{8}{l}{$^{(b)}$ replacement for flared observation 0106060801}\\
\multicolumn{8}{l}{$^{(c)}$ background pointing (see text)}
\label{tab:obslog}
\end{tabular}
\end{center}
\end{table*}

\subsection{Observations}
\label{subs:obs}

\begin{figure}
\centering
\includegraphics[width=0.45\textwidth,bb=0 0 37 44]{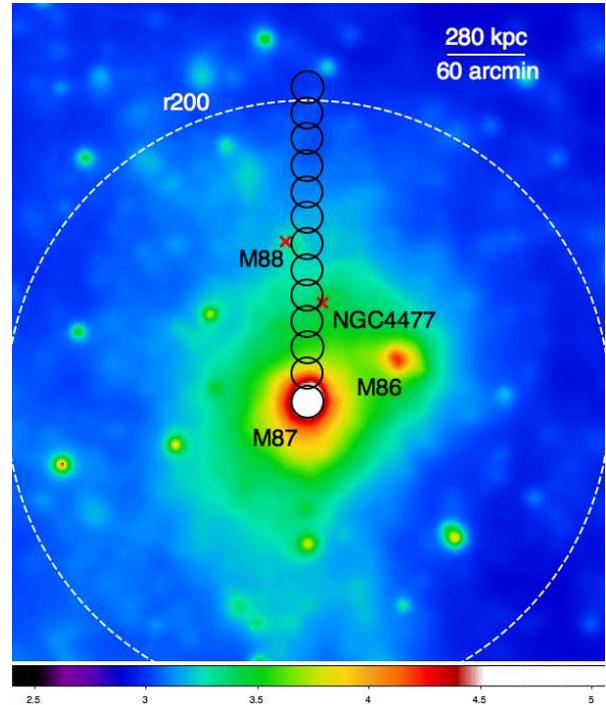}
\caption{Adaptively smoothed \rosat\ PSPC image of the Virgo Cluster in the 0.5--2.0~keV band with the \xmm\ pointing positions over-plotted as 12\arcmin\ radius circles. The color table corresponds to a logarithmic brightness scale. The background pointing lies 1.5$^{\circ}$ to the north of the northernmost pointing shown. M87 is the brightest cluster galaxy and sits in the centre of the Virgo Cluster, M86 is in the centre of an X-ray bright a sub-clump. The large elliptical galaxy NGC4477 and the spiral galaxy M88 are likely associated with sub-clusters  and are marked with red crosses. The approximate $r_{200}$ of the Virgo Cluster is shown with a dashed circle. }
\label{fig:rosat}
\end{figure}

\begin{figure*}
\centering
\includegraphics[width=0.95\textwidth,bb=14 14 435 555]{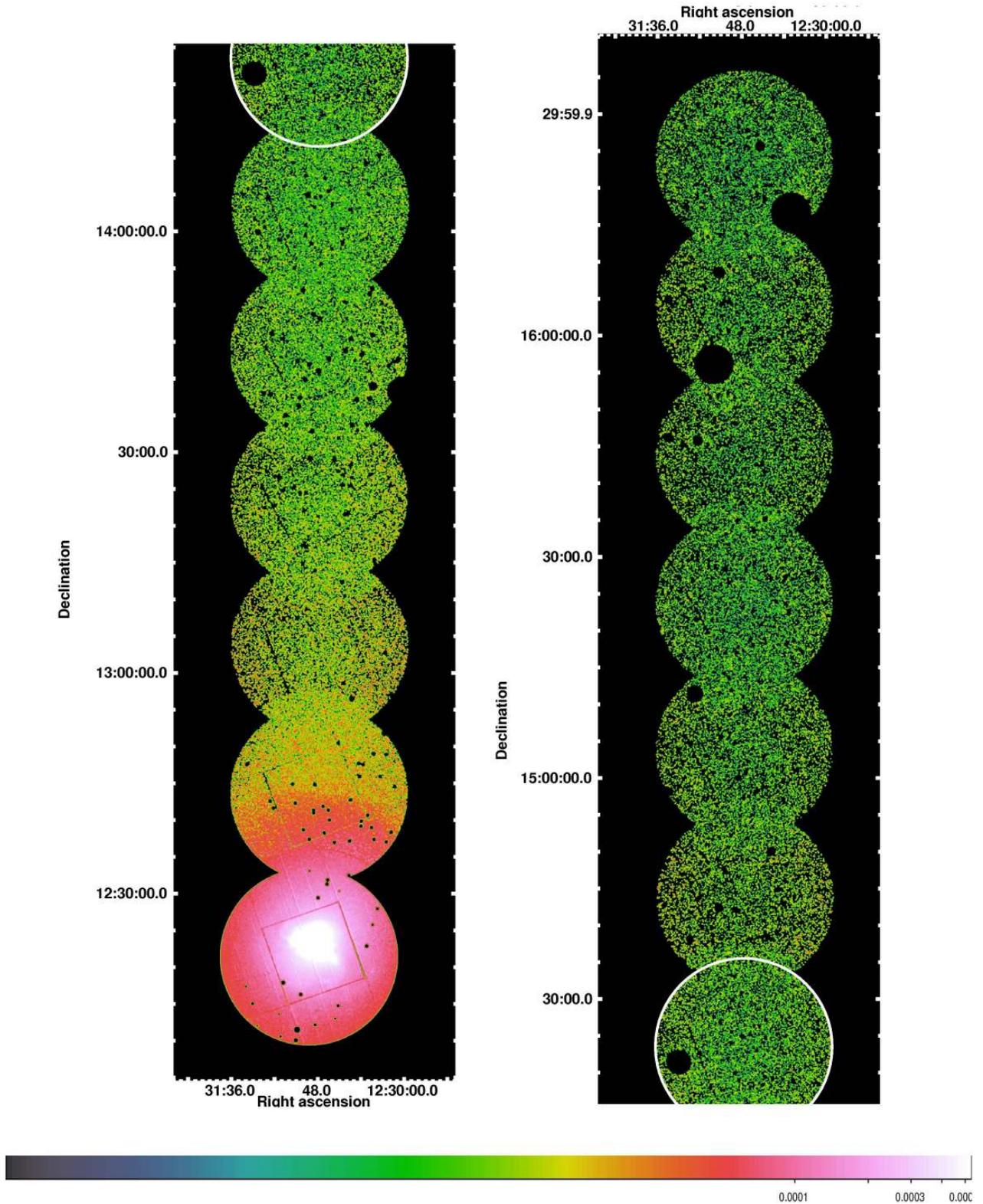}
\caption{Mosaic \xmm\ image in the $0.6-3.0\,\text{keV}$ energy range. The background pointing lies 1.5$^{\circ}$ to the north of the northernmost pointing shown. Only the central region within the radius of 12\arcmin\ from the telescope's aim-point was used for each instrument. The image has been exposure-corrected, smoothed with a gaussian with the width of 3 pixels, and the point sources have been removed. The left panel shows the southern part of the mosaic and the right panel shows the northern part. The white circle marks the overlap in both panels. The central part of the cluster is clearly dominated by the emission of M87, the cold front discussed in \citet{2010arXiv1002.0395S} is visible at $\sim20$\arcmin\ to the north of the center.}
\label{fig:panorama}
\end{figure*}

A total of fourteen \xmm\ observations were used, the details of which are listed in Table~\ref{tab:obslog}.  The observations were carried out in June and July 2002, with the exception of the southernmost ones. Thirteen of the pointings are partially overlapping and cover a stripe in the sky with dimensions roughly 30\arcmin\ (East-West) by 4.5$^{\circ}$ (North-South). The combined image of these pointings using the MOS detectors is shown in Fig.~\ref{fig:panorama}. The northernmost observation located $\sim5.5^{\circ}$ from the cluster center (not shown in Fig.~\ref{fig:rosat} and \ref{fig:panorama}) lies beyond the virial radius of Virgo, contains no significant ICM emission, and was used to determine the local X-ray background.

\subsection{Data analysis}
\label{subs:analysis}

The data analysis was carried out using the \xmm\ Science Analysis System (SAS) version 9.0.0. The filtering and extraction of the MOS data products and the subtraction of their instrumental background was performed using the \xmm\ \emph{Extended Source Analysis Software} ({\tt XMM-ESAS}) and methods described in \citet{kuntz2008,snowden2007,snowden2010}. The filtering and subtraction of the instrumental background for the pn data was based on that performed by \citet{werner2008}.

In brief, to remove the soft proton flares, we used the {\tt mos-filter} procedure in {\tt XMM-ESAS} \citep{snowden2010}. The MOS detector background was modeled and subtracted using the filter-wheel-closed (FWC) data together with data from the unexposed corners of the CCDs. The total count rates and hardness ratios are determined in the unexposed regions of the CCDs for each source observation and a data base of all archived observations is searched for FWC data with similar count rates and hardness ratios. FWC data with the most similar count rates and hardness ratios are then scaled by the ratio of the source observation spectra from the CCD corners to the FWC spectra from the corners. These scaled FWC data are then used to produce the background files for the appropriate regions. The background modeling is done for each MOS chip individually. 

To remove the soft proton flares from the EPIC/pn detectors, we extracted light curves in the hard energy band, 10--14 keV, and created count rate histograms which have a roughly Gaussian peak at the nominal count rate. We excluded the time periods when the count rate exceeded the mean by more than 3$\sigma$. We repeated this procedure in the soft energy, 0.3--1~keV, band. The spectral properties of the EPIC/pn detector background are relatively stable and the chip-to-chip variations are not significant. For EPIC/pn, we therefore used a stacked FWC data set which we again scaled by the ratio of the number of the counts in the unexposed corners of the source observation and the FWC data. We also tried to normalize by the ratio of the observed and FWC count rates in the 10--14~keV band, where practically all counts are instrumental and the number of source counts is negligible. Both methods of scaling produced consistent results.

Table~\ref{tab:obslog} shows the observation ID, date, revolution number, exposure times before and after cleaning for each instrument, and the coordinates of each pointing. For both EPIC/MOS and EPIC/pn, we check for residual soft proton contamination by comparing the 10--12~keV (10--14~keV for EPIC/pn) count rate per unit area within the field of view and in the unexposed corners. Soft protons are focused by the mirrors of the telescopes and they do not affect the unexposed areas of the detectors. Therefore, this comparison, in the hard energy band where the number of source counts is negligible, is a good way to detect a significant residual soft proton contamination \citep[see][]{leccardi2008}.
Two of the observations (obs. ID 0106060701 and 0106061101) were affected strongly by flares and were not used in the spectral analysis due to residual soft-proton contamination. For observation ID 0106060201, the useful pn exposure time after cleaning was only 1.7~ks and the comparison of the exposed and unexposed areas of the detector in the hard energy band indicates residual soft proton contamination. Therefore, it too has not been used in the further analysis. 

Images were extracted in the $0.6-3.0\,\text{keV}$ energy range.
We stacked the images for all three instruments and used the SAS task \texttt{ewavelet} to detect point sources down to $5\sigma$ significance. The stacked images were visually inspected to correct for false detections, sources with larger dimensions, and point sources that were missed by \texttt{ewavelet}. For the central pointing, we used the list of point sources from an earlier analysis by \citet{2010arXiv1002.0395S}. For a $z=0.98$ background cluster located about 1.3$^{\circ}$ north of M87  \citep{fassbender2010}, we excluded a region within its $r_{500}$. A total of 580 point sources were excluded from the subsequent analysis.

To minimize systematic uncertainties associated with measurements at large off-axis angles (e.g. effective area calibration uncertainties), we have only used the data collected from within circular extraction regions of radius 12\arcmin, centered on the aim points of the individual telescopes for each pointing. 

\subsection{Spectral analysis}
\label{subs:spectral_analysis}

We extracted spectra from all clean observations north of the central pointing. 
Previous results from these data for the cluster's central region are discussed by e.g. \citet{belsole2001m87,molendi2001m87,boehringer2001m87,matsushita2002m87a,matsushita2003m87a,werner2006b,simionescu2007,simionescu2008a,2010arXiv1002.0395S}.

The spectra were extracted from a set of annular regions centered on the active nucleus of M87 ($\alpha=12\text{h}\thinspace30\text{m}\thinspace49.4\text{s}$, $\delta=12^{\circ}\thinspace23'\thinspace28''$). The widths of the annuli were varied to maintain a nearly constant number of instrumental background subtracted counts (10000) in each region. The SAS task \texttt{rmfgen} was used to create response matrices (RMF). Auxiliary response files (ARF) were created for each region using the task \texttt{arfgen}. 

We used {\tt XSPEC 12.6.0} \citep{arnaud1996} to model the spectra. Fits were carried out in the $0.4-7.0\,\text{keV}$ energy range using the combined data from all three detectors. We excluded the $1.2-1.8\,\text{keV}$ interval due to the instrumental Al and Si lines.  For the diffuse X-ray emission, we used a single-temperature \texttt{apec} model \citep{smith2001}, which describes an optically thin plasma in collisional ionization equilibrium. The redshift was fixed to that of M87 ($z=4.36\times10^{-3}$). For each pointing, we fixed the Galactic absorption column density to the value determined by the Leiden/Argentine/Bonn Survey \citep{kalberla2005}. The metallicities  are reported relative to the Solar abundances given by \citet{grevesse1998}. We used the extended C-statistic (which allows for background subtraction) in all the spectral modeling. 

\subsection{X-ray foreground and background}

\begin{table}
\setlength{\extrarowheight}{4pt}
\begin{center}
\caption{Fore- and background model parameters. Fluxes are given in the 2--10~keV interval. The second column lists the photon index of the absorbed power-law component and the temperatures of the two foreground thermal components. GH stands for Galactic Halo, LHB for Local Hot Bubble.}
\begin{tabular}{l|cc}
\hline\hline
component              & $\Gamma$/Temperature [keV] & flux $\left[\text{erg}\thinspace\text{s}^{-1}\thinspace\text{cm}^{-2}\thinspace\text{deg}^{-2}\right]$ \\
\hline
power-law              & $1.44\pm0.03$           & $11.90_{-0.07}^{+0.06}\times10^{-11}$\\
apec$_{\text{GH}}$ & $0.191_{-0.004}^{+0.005}$  & $\left(6.32\pm0.32\right)\times10^{-16}$ \\
apec$_{\text{LHB}}$ & $0.08^{\dagger}$           & $8.47_{-0.64}^{+0.72}\times10^{-21}$ \\
\hline
\multicolumn{2}{l}{$^{\dagger}$ fixed value}\\
\end{tabular}
\label{tab:backgr}
\end{center}
\end{table}

We extracted the cosmic X-ray foreground/background (CXFB) spectrum from the northernmost pointing using the combined data from all three detectors, after removing the point sources. We assumed a sum of three components in the CXFB model -- an absorbed power-law from the unresolved point sources \citep{deluca2004}, an $\sim$0.2~keV emission from the Galactic halo \citep{kuntz2000} and the Local Hot Bubble emission modeled by 0.08~keV thermal plasma \citep[LHB\footnote{The LHB emission is produced within the Galactic absorption column and therefore this component is unabsorbed.},][]{1996A&A...305..308S,kuntz2000}. During the CXFB model fitting, the normalizations of the three components, the power-law index, and the temperature of the Galactic halo gas were left to vary independently. The metallicities of the foreground thermal plasmas were fixed to the Solar value. The absorption column density was fixed at $N_{\mathrm{H}}=2.13\times10^{20}\thinspace\text{cm}^{-2}$ \citep{kalberla2005}. The resulting foreground/background parameters are summarized in Table~\ref{tab:backgr}. The observed flux of the power-law component is lower than the value reported by \citet{deluca2004}, which is to be expected given that the point sources detected and resolved in our analysis were excluded.

All spectra used this same CXFB model, with the normalizations scaled to account for the areas of the extraction regions.

The north-polar spur region, with its elevated soft foreground emission, is relatively strong throughout the eastern parts of the Virgo Cluster. However, our pointings are in a direction that avoids this emission. 

\section{Results}
\label{sect:results}

\subsection{X-ray surface brightness}

\begin{figure}
\includegraphics[angle=0,width=.47\textwidth,bb=18 144 592 718]{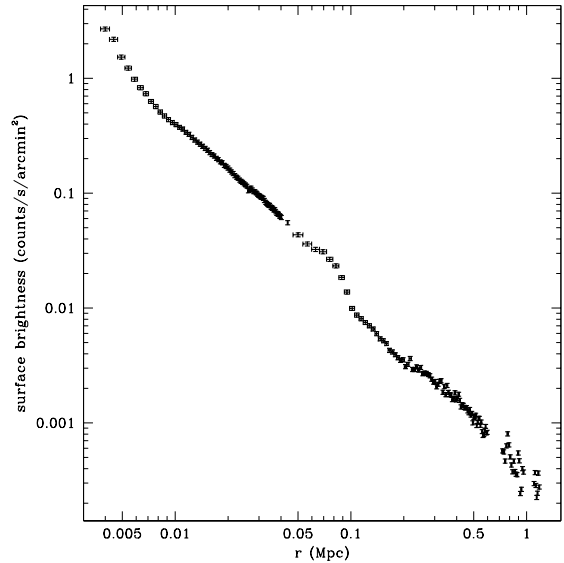}
\caption{Background-subtracted surface brightness profile in the 0.6--3.0~keV band. The gaps in the profile indicate missing data due to flared observations. The virial radius, which we define as $r_{200}$, lies at about 1.08~Mpc.}
\label{fig:brig_profile}
\end{figure}

In Fig.~\ref{fig:brig_profile} we present the background-subtracted surface brightness profile of the Virgo Cluster obtained from the combined, exposure-corrected MOS images in the $0.6-3.0\,\text{keV}$ band. It was extracted from 199 annular regions stretching from the center out to $\sim255$\arcmin\ (projected distance $\sim1.2\thinspace\text{Mpc}$). In the cluster center, out to $\sim8.5$\arcmin\ $=40\thinspace\text{kpc}$, the regions are 0.1\arcmin\ wide; at larger intermediate radii their width is $\sim$1.35\arcmin, and beyond 700~kpc the binsize increases to 2.7\arcmin. The gaps in the profile indicate missing data due to flared observations. The background was assumed constant over the whole cluster, with a value equal to the average surface brightness of the outer background pointing. 

The profile shows a prominent discontinuity associated with a cold front at 90 kpc. This feature has previously been discussed by \citet{2010arXiv1002.0395S}. Outside the cold front, the surface brightness distribution can be described by a power-law model $S_{\mathrm{X}}\propto r^{-\alpha}$ with index $\alpha=1.34\pm0.01$. We verified that there is no significant difference between the slopes of surface brightness profiles obtained for different low-energy bands. 

In the faint outskirts, beyond $r\sim700$~kpc, the surface brightness starts to fluctuate significantly and the profile becomes more uneven. We note that the individual surface brightness bins in these outer regions contain at least $\sim$500 counts. The fluctuations are therefore formally highly statistically significant.

\subsection{Spectral results}
\label{specres}

\begin{figure}
\includegraphics[angle=0,width=.50\textwidth,bb=18 144 592 718]{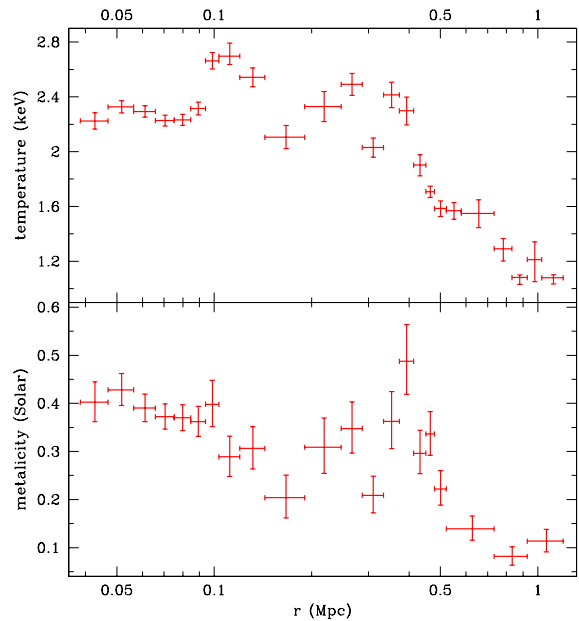}
\caption{Projected temperature and metallicity profiles. The virial radius, which we define as $r_{200}$, lies at about 1.08~Mpc.}
\label{fig:tempmetal}
\end{figure}

\begin{figure*}
\begin{minipage}{0.48\textwidth}
\includegraphics[angle=270,width=.9\textwidth,angle=90]{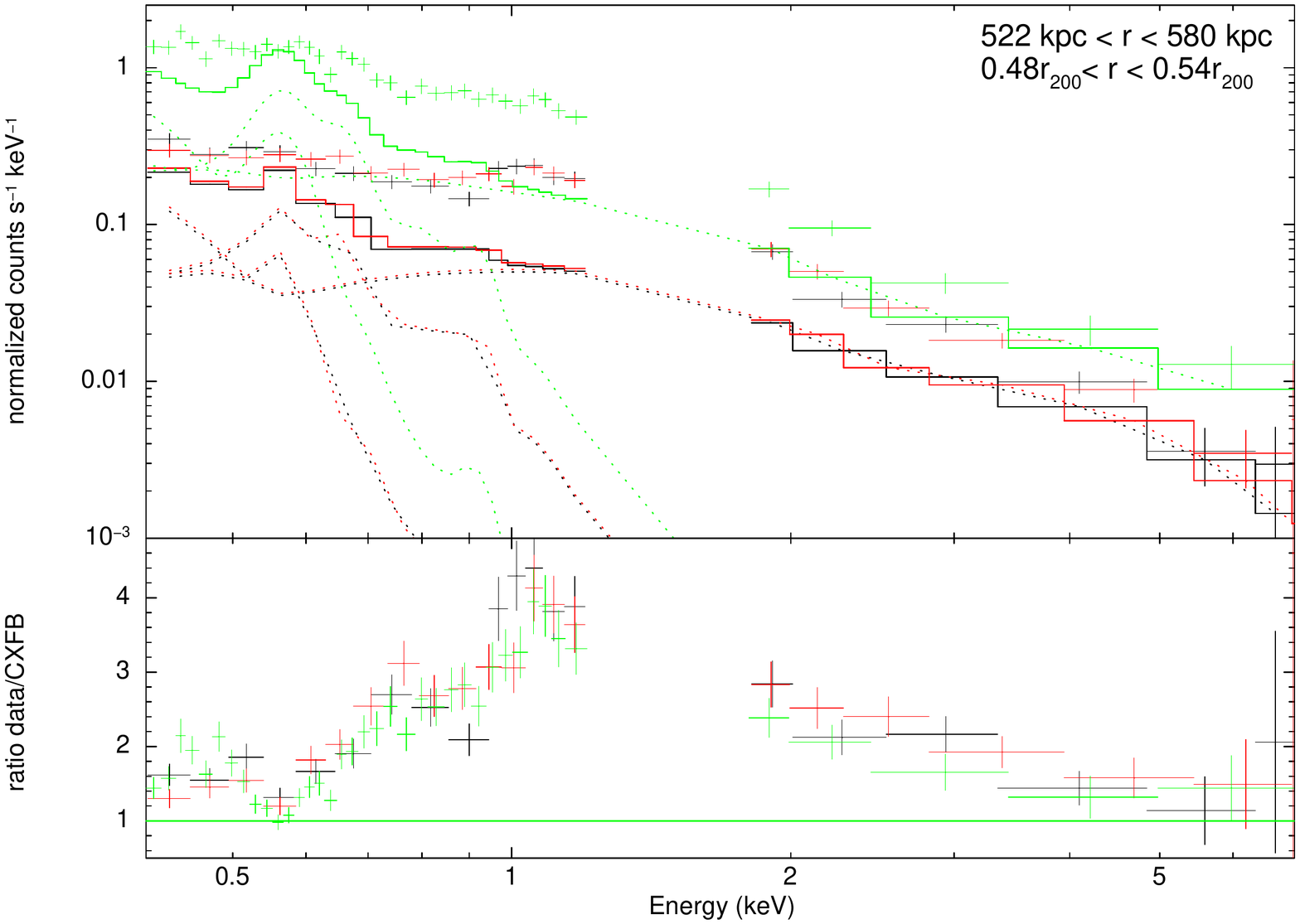}
\includegraphics[angle=270,width=.9\textwidth,angle=90]{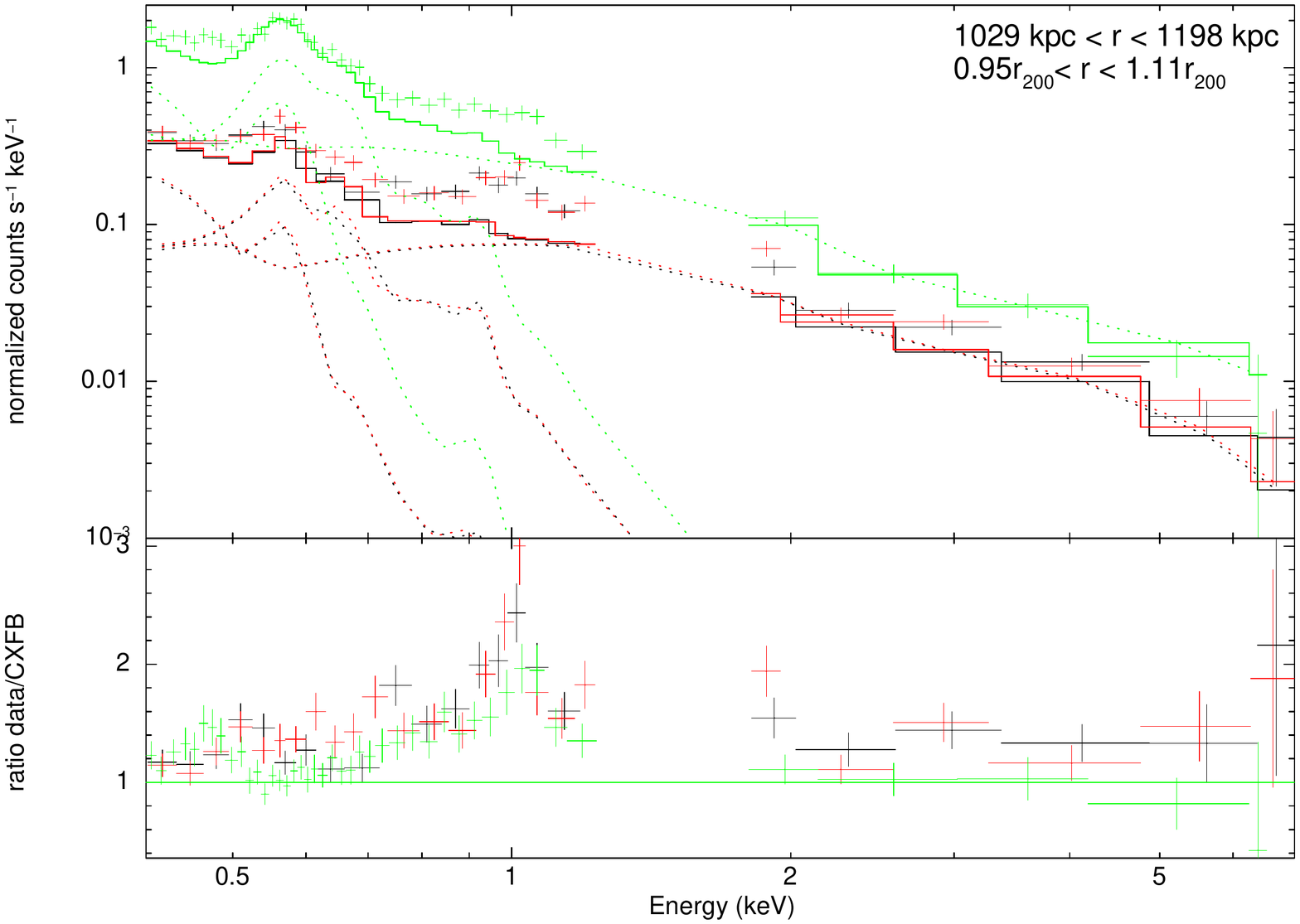}
\end{minipage}
\begin{minipage}{0.48\textwidth}
\includegraphics[angle=270,width=.9\textwidth,angle=90]{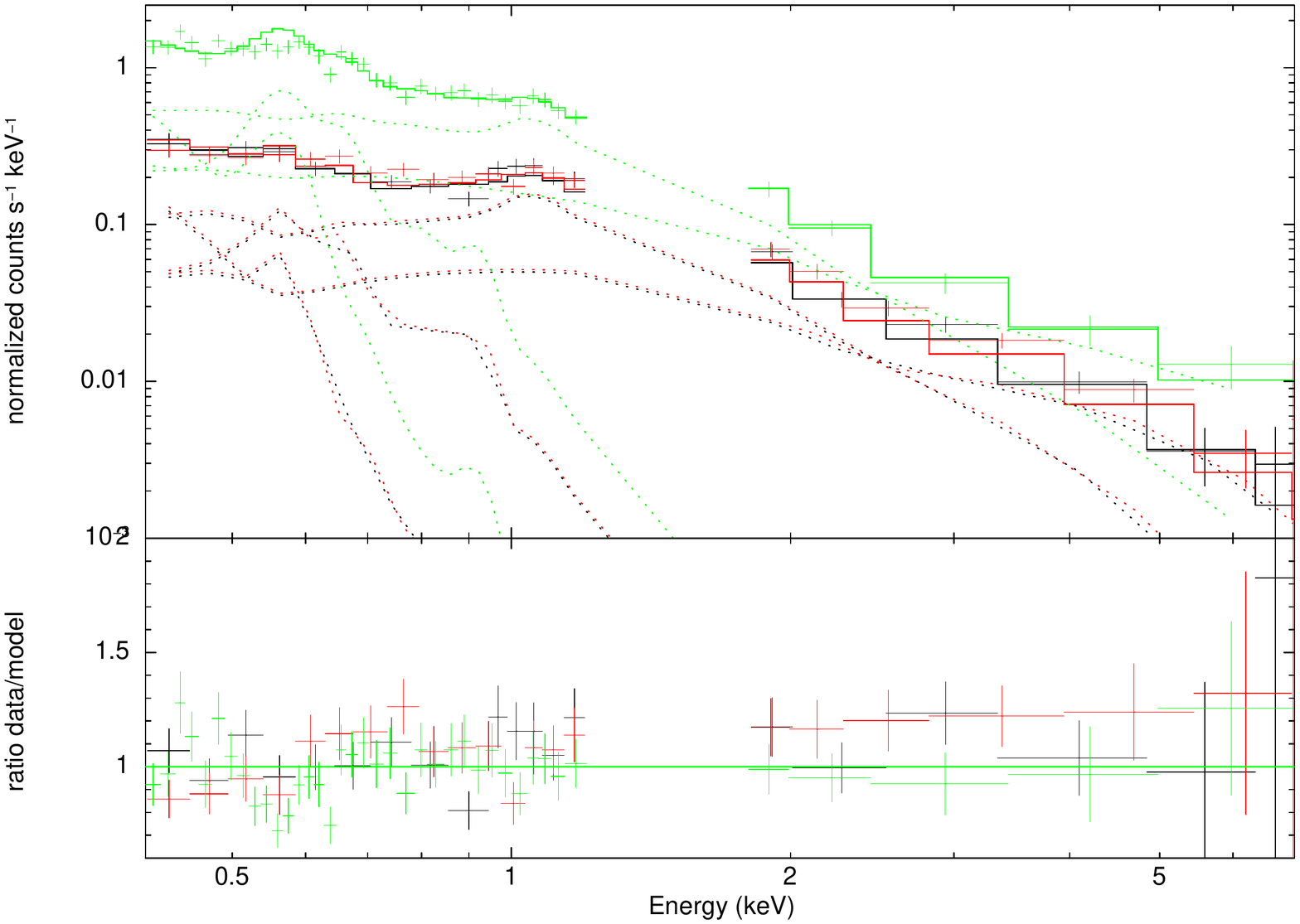}
\includegraphics[angle=270,width=.9\textwidth,angle=90]{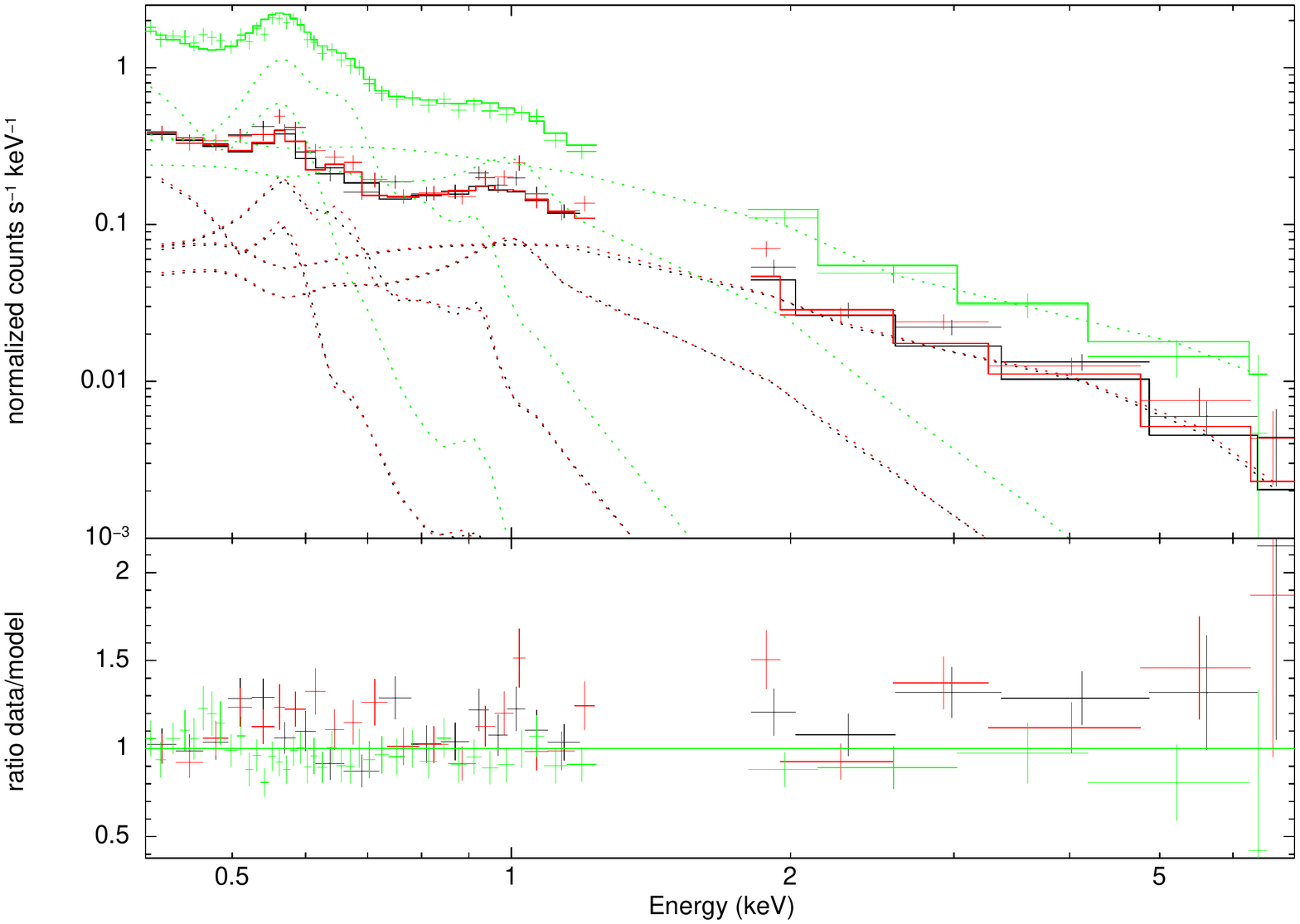}
\end{minipage}
\caption{Examples of the analyzed spectra from $\sim$0.5$r_{200}$ and from $\sim$$r_{200}$. Data from all three EPIC instruments -- MOS~1 (black), MOS~2 (red) and pn (green) -- were fitted simultaneously. \emph{Left:} Data with over-plotted CXFB model. Bottom part of each plot shows the ratio of the total to CXFB flux, thus indicating the amount of cluster emission. \emph{Right:} Spectra with over-plotted best fit model, which includes the cluster emission.}
\label{fig:spectra_backgr}
\end{figure*}

\begin{figure}
\includegraphics[angle=0,width=.5\textwidth,bb=18 144 592 718]{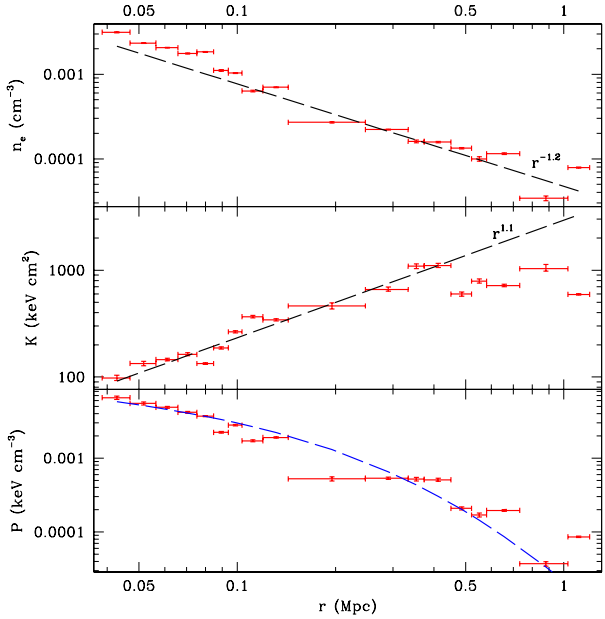}
\caption{The deprojected electron density ($n_{\mathrm{e}}$), entropy ($K$), and pressure ($P$) profiles. For the density profile, we overlay the best fit power-law model, $n_{\mathrm{e}}\propto r^{-1.2}$. On the entropy profile we over-plot $K\propto r^{1.1}$, which is expected for gravitationally collapsed gas in hydrostatic equilibrium. On the pressure profile, we over-plot the extrapolated average profile determined for a sample of clusters within $\sim$0.5$r_{200}$ by \citet{arnaud2010}. The virial radius, which we define as $r_{200}$, lies at about 1.08~Mpc.}
\label{fig:nesp}
\end{figure}

\begin{table}
\setlength{\extrarowheight}{4pt}
\begin{center}
\caption{Projected temperatures, metallicities, and normalizations obtained from spectral fitting. The first and the second columns give the inner and outer radii of the annular extraction regions.}
\begin{tabular}{ccccc}
\hline\hline
\multicolumn{2}{c}{Radius [arcmin]} & Temperature & Metallicity & Normalization \\
Inner & Outer & [keV] & [Solar] & $\left[1\times10^{-5}\thinspace\text{arcmin}^{-2}\right]$\\
\hline
8.2 & 10 & $2.224_{-0.060}^{+0.061}$ & $0.40_{-0.04}^{+0.04}$ & $121.39_{-2.73}^{+2.72}$ \\
10 & 12 & $2.326_{-0.045}^{+0.045}$ & $0.43_{-0.03}^{+0.03}$ & $91.63_{-1.54 }^{+1.54}$ \\
12 & 14 & $2.293_{-0.041}^{+0.042}$ & $0.39_{-0.03}^{+0.03}$ & $74.46_{-1.13}^{+1.14}$\\
14 & 16 & $2.227_{-0.039}^{+0.040}$ & $0.37_{-0.03}^{+0.03}$ & $60.48_{-0.88}^{+0.89}$\\
16 & 18 & $2.232_{-0.040}^{+0.041}$ & $0.37_{-0.03}^{+0.03}$ & $47.68_{-0.72}^{+0.72}$\\
18 & 20 & $2.314_{-0.047}^{+0.047}$ & $0.36_{-0.03}^{+0.03}$ & $35.22_{-0.59}^{+0.59}$\\
20 & 22 & $2.663_{-0.061}^{+0.061}$ & $0.40_{-0.05}^{+0.05}$ & $22.25_{-0.52}^{+0.53}$\\
22 & 25 & $2.697_{-0.060}^{+0.096}$ & $0.29_{-0.04}^{+0.04}$ & $16.53_{-0.32}^{+0.32}$\\
25 & 30 & $2.542_{-0.068}^{+0.069}$ & $0.31_{-0.04}^{+0.05}$ & $13.43_{-0.30}^{+0.30}$\\
30 & 40 & $2.105_{-0.083}^{+0.086}$ & $0.20_{-0.04}^{+0.05}$ & $10.10_{-0.33}^{+0.33}$\\
40 & 52 & $2.328_{-0.109}^{+0.111}$ & $0.31_{-0.05}^{+0.06}$ & $7.01_{-0.23}^{+0.23}$\\
52 & 61 & $2.491_{-0.080}^{+0.081}$ & $0.35_{-0.05}^{+0.06}$ & $5.64_{-0.15}^{+0.15}$\\
61 & 71 & $2.030_{-0.070}^{+0.070}$ & $0.21_{-0.04}^{+0.04}$ & $4.46_{-0.13}^{+0.13}$\\
71 & 79 & $2.413_{-0.093}^{+0.093}$ & $0.36_{-0.06}^{+0.06}$ & $3.93_{-0.12}^{+0.13}$\\
79 & 88 & $2.298_{-0.102}^{+0.101}$ & $0.49_{-0.07}^{+0.08}$ & $3.00_{-0.11}^{+0.11}$\\
88 & 96 & $1.902_{-0.078}^{+0.077}$ & $0.30_{-0.04}^{+0.05}$ & $2.99_{-0.10}^{+0.10}$\\
96 & 102 & $1.708_{-0.042}^{+0.039}$ & $0.34_{-0.04}^{+0.05}$ & $2.48_{-0.09}^{+0.09}$\\
102 & 111 & $1.585_{-0.058}^{+0.055}$ & $0.22_{-0.03}^{+0.04}$ & $2.36_{-0.09}^{+0.09}$ \\
111 & 123 & $1.569_{-0.062}^{+0.061}$ & \multirow{2}{*}{$0.14_{-0.02}^{+0.03}$} & $2.16_{-0.07}^{+0.08}$\\
123 & 156 & $1.549_{-0.103}^{+0.098}$ & & $1.64_{-0.07}^{+0.07}$ \\
156 & 177 & $1.292_{-0.090}^{+0.074}$ & \multirow{2}{*}{$0.08_{-0.02}^{+0.02}$} & $1.47_{-0.07}^{+0.08}$ \\
177 & 197 & $1.081_{-0.050}^{+0.020}$ & & $0.87_{-0.06}^{+0.06}$ \\
197 & 219 & $1.213_{-0.161}^{+0.129}$ & \multirow{2}{*}{$0.11_{-0.02}^{+0.02}$} & $1.07_{-0.09}^{+0.09}$ \\
219 & 255 & $1.080_{-0.045}^{+0.022}$ & & $0.64_{-0.05}^{+0.05}$ \\
\hline
\label{tab:spectra}
\end{tabular}
\end{center}
\end{table}

The results of our spectral fits are shown in Table~\ref{tab:spectra} and in Fig.~\ref{fig:tempmetal}. Because we fitted the data using C-statistics, we do not have have a goodness of fit indicator similar to the reduced $\chi^2$. In order to confirm that the fits are acceptable, we re-binned the data to a minimum of 50 counts per bin and evaluated our best fit model using $\chi^2$ statistics. All the reduced $\chi^2$ values for our best fit models are in the range of 0.86--1.38. These values show that, given the present systematics, our fits are acceptable. The goodness of fit estimated from the $\chi^2$ statistics does not show a systematic trend with radius: the fits at large radii are, on average, as good as the fits at small radii.

Both the temperature and metallicity profiles show discontinuities at the cold front ($r\sim90$~kpc), previously studied by \citet{2010arXiv1002.0395S} and modeled by \citet{roediger2010}. This feature can be explained by the sloshing of the dense gas associated with the cooling core in the gravitational potential of the cluster.

Beyond a radius $r\sim450$~kpc, the temperature and metallicity both drop abruptly. At the virial radius, the measured temperature reaches $kT=1.08^{+0.02}_{-0.05}$~keV and the metallicity is $Z=0.11\pm0.02$~Solar. This is the first time that the metallicity at $r_{200}$ has been measured with high ($>$5.5$\sigma$) statistical significance. This is possible in Virgo due to its proximity and the low ICM temperature, for which the Fe-L line emission is strong. At $\sim$1~keV, where the Fe-L emission peaks, the effective area of the \xmm\ mirrors is large and the signal-to-detector background contrast is optimal. 

The strength of the Fe-L signal can be seen in Fig.~\ref{fig:spectra_backgr}, where we show the spectra from the radius 522--580~kpc or $r\sim0.5\, r_{200}$ and for our outermost source extraction region at 1029--1198~kpc, or $r\sim r_{200}$. The solid lines on the left show the background models for the given spectra; the bottom panels show the ratio of the measured X-ray flux to the CXFB flux. We can see clear excess signal above the background, representing the cluster emission. A large part of the detected signal from near the cluster's virial radius arises in the 0.8--1.2~keV range, and is associated with Fe-L line emission. The right panels show the same data with the over-plotted best fit model including the cluster emission. 

Deprojected profiles of the temperature and density were derived from the combined set of spectra using the \texttt{projct} model in {\tt XSPEC}. In order to ensure the stability of the fit, we linked some of the parameters in neighboring annular regions. The deprojected density profile is shown in the top panel of Fig.~\ref{fig:nesp}. The measured electron density approximately follows a power-law model $n_{\mathrm{e}}\propto r^{-\beta}$ with index $\beta=1.21\pm0.12$.  

Based on the deprojected values of the temperature and density, we also calculated the entropy $\left(K=\frac{kT}{n_{\mathrm{e}}^{2/3}}\right)$ and pressure $\left(p=n_{\mathrm{e}}kT\right)$ profiles. These are also shown in Fig.~\ref{fig:nesp}. On the entropy profile we over-plot $K\propto r^{1.1}$, which is expected for gravitationally collapsed gas in hydrostatic equilibrium \citep{tozzi2001,2005RvMP...77..207V}. Out to $r\sim450$~kpc, the entropy profile follows the expected power-law shape, but further out it becomes flatter, staying consistently below the expected value by a factor of 2--2.5. On the pressure profile, we over-plot the average profile for a sample of clusters observed within $\sim$0.5$r_{200}$ \citep{arnaud2010} extrapolated to larger radii. Our measured profile shows clear departures from this average shape: a deficit is observed at $\sim$200~kpc, and enhancements are seen at $\sim$400~kpc and $\sim$600~kpc. Because the Virgo Cluster is relatively unrelaxed, the results of the deprojection analysis, which assumes spherical symmetry, should be treated with caution. The systematic uncertainties due to the departures from spherical symmetry are significantly larger than the measurement errors. 

The last data point in the deprojected electron density profile is most likely overestimated due to unaccounted emission from outside of this radius, biasing the pressure in the last radial bin high and the entropy low. A small ringing effect from this artifact will also be projected onto the next bin inwards. Such artifacts do not affect our main results on the density and entropy profiles. 
The observed pressure enhancement at 400~kpc coincides with a region of increased X-ray surface brightness, a possible sub-group surrounding the massive elliptical galaxy NGC4477, lying just outside of our \xmm\ extraction region (indicated in Fig.~\ref{fig:rosat}). The pressure increase at 600~kpc coincides with an X-ray bright region surrounding the massive spiral galaxy M88 (see Fig.~\ref{fig:rosat}). Projection effects associated with the gas of these sub-groups could be responsible for the observed features in the pressure profile. 

The flattening of the entropy profile is robustly measured and particularly interesting. The observed entropy beyond the radius of $\sim$450~kpc is consistently a factor of 2--2.5 lower than the expected value, implying that the observed density is higher and/or the observed temperature is lower than expected.

\subsection{Systematic errors}
\label{sect:systematics}

\begin{figure*}
\begin{minipage}{0.48\textwidth}
\includegraphics[angle=0,width=\textwidth,bb=18 144 592 718]{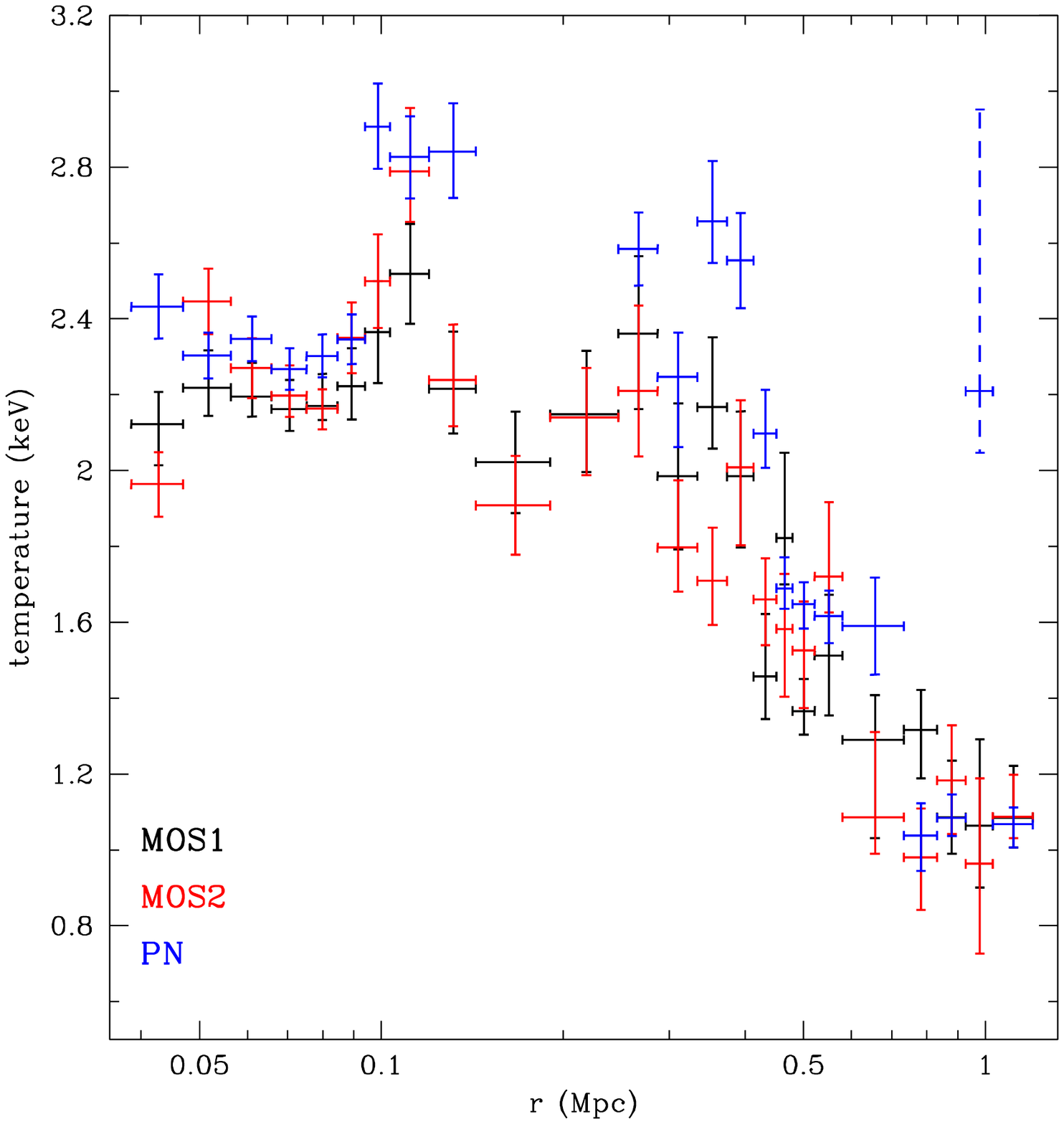}
\end{minipage}
\begin{minipage}{0.48\textwidth}
\includegraphics[angle=0,width=\textwidth,bb=18 144 592 718]{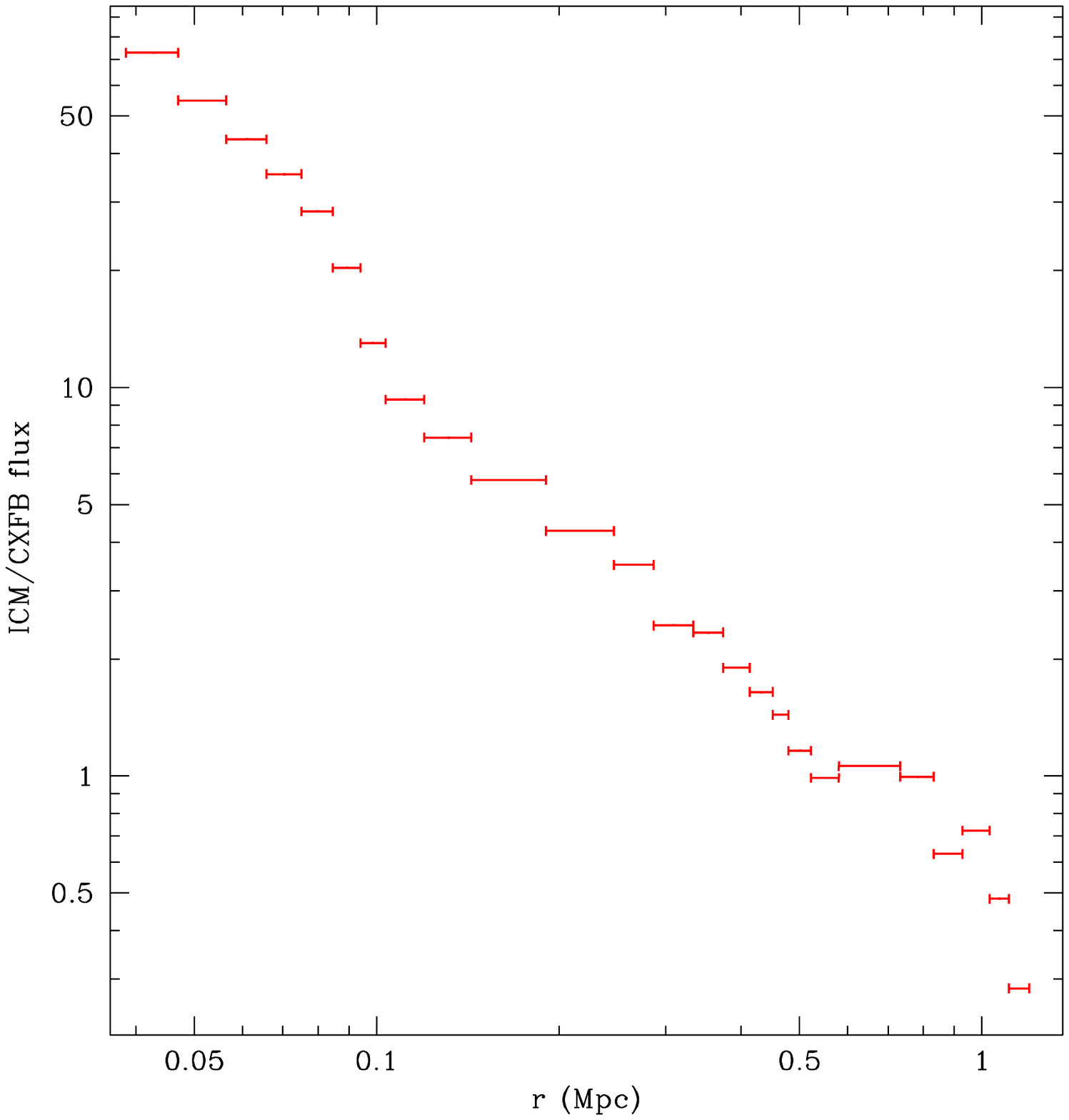}
\end{minipage}
\caption{{\it Left panel: }Temperature profiles for the separate EPIC instruments -- MOS~1 (black), MOS~2 (red) and pn (blue). EPIC/pn was not used in the analysis at the radius at which the pn value is shown with dashed lines. {\it Right panel: }Ratio of the flux of the ICM versus the flux of the CXFB for the annular regions. Fluxes were determined in the $0.4-3\thinspace\text{keV}$ range. }
\label{fig:single_flux}
\end{figure*}

\begin{figure}
\begin{minipage}{0.50\textwidth}
\includegraphics[angle=0,width=\textwidth,bb=18 144 592 718]{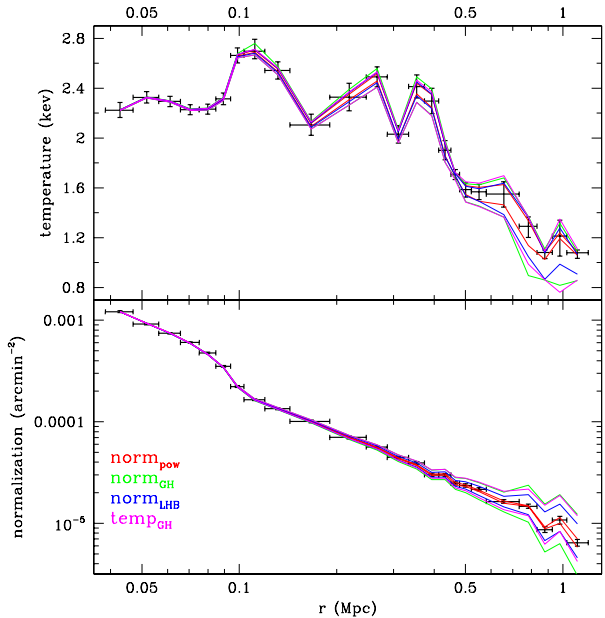}
\end{minipage}
\caption{Projected temperatures (upper panel) and normalizations (lower panel) with over-plotted profiles obtained by varying CXB parameters (see the text for details). We show the bracketing values due to the uncertainties on the normalization of the power-law background component ($\text{norm}_{\text{pow}}$), normalization of the Galactic halo emission ($\text{norm}_{\text{GH}}$), normalization of the LHB emission component ($\text{norm}_{\text{LHB}}$), and the temperature of the Galactic halo emission component ($\text{temp}_{\text{GH}}$).}
\label{fig:sigma_profile}
\end{figure}

\begin{figure}
\begin{minipage}{0.50\textwidth}
\includegraphics[angle=0,width=\textwidth,bb=18 144 592 718]{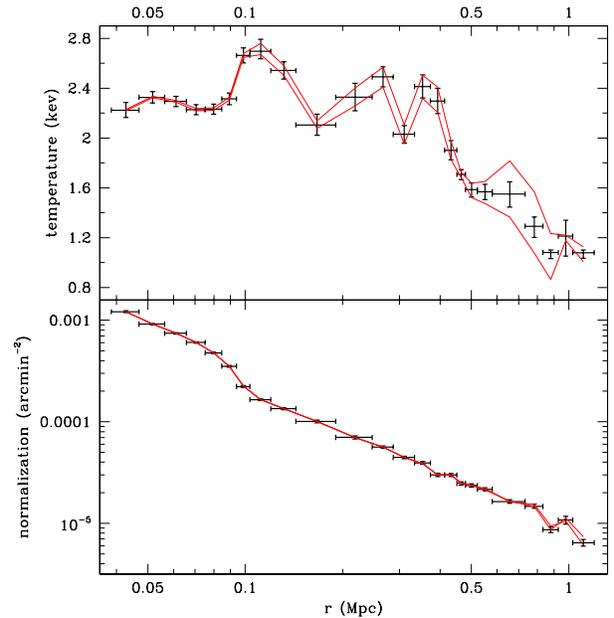}
\end{minipage}
\caption{Projected temperatures (upper panel) and normalizations (lower panel) with over-plotted profiles obtained by varying the instrumental background (see the text for details).}
\label{fig:inst_profile}
\end{figure}

In order to test the robustness of our results, we fitted the data obtained by the three \xmm\ EPIC detectors separately. The temperature results are shown in the left panel of Fig.~\ref{fig:single_flux}. While the results obtained by the MOS1 and MOS2 detectors agree remarkably well at almost all radii, the temperature values obtained by EPIC/pn are systematically higher. The largest inconsistency between the MOS and pn detectors is in the one but last annulus (shown with dashed lines in Fig.~\ref{fig:single_flux}). We can find no reason for this inconsistency - it appears not to be due to residual soft proton contamination or technical issues. On the grounds of their gross inconsistency with the two independent MOS detectors (which are fully consistent with each other), at this radius we have excluded the EPIC/pn data from the analysis. With the EPIC/pn data included, the best fit temperature in this region is $1.46\pm0.17$~keV.

However, we emphasize that the results are in a relatively good agreement at most radii and the general trend with the temperature decreasing at large radii is robust.

\subsubsection{Cosmic X-ray fore- and background}

The right panel of Fig.~\ref{fig:single_flux} shows the radial dependence of the ratio of the ICM/CXFB fluxes in the $0.4-3.0\,\text{keV}$ range. The ratio drops below unity beyond $\sim800\thinspace\text{kpc}$. However, as we show in the left-hand side of Fig.~\ref{fig:spectra_backgr}, the signal at $\sim$1~keV is relatively strong even in the outermost annulus.

We determine the variation of the soft foreground components within the area covered by our pointings using the images obtained in the \rosat\ All Sky Survey (RASS) in the soft 0.1--0.3 keV (R1+R2) band. Most of the LHB and GH foreground is emitted in this energy band. Studying the point-to-point variation in these \rosat\ all-sky maps we determined a conservative upper limit of 30 per cent for the variation of the LHB and Galactic halo fluxes along our investigated region. The Galactic halo temperature was found using the \rosat\ hardness ratio maps to vary by up to 20 per cent of its value. Due to the exclusion of point sources down to a faint level, the root mean square variation of the normalization of the power-law component is expected to be smaller than the statistical uncertainty with which its value has been determined in the background pointing. Thus we varied this parameter by its 3$\sigma$ statistical error. 

These uncertainties were used to determine the bracketing values for the temperature and normalization profiles shown in Fig.~\ref{fig:sigma_profile}. The background parameters were up- and downscaled one at a time, leaving the others with their original values, and the model was refitted. Although at radii larger than $\sim$500~kpc the systematic errors dominate over the statistical uncertainties, the values remain well determined and the trends in the profiles are robust.  

\subsubsection{Instrumental background}

The uncertainty in the normalization of the instrumental background is 3--5 per cent. These uncertainties arise mainly from the Poisson error associated with the scaling factor which is used to account for the variation of the instrumental background level with time. The scaling factors of all three detectors were simultaneously raised or lowered by $1\sigma$ to achieve a conservative estimate. 
The bracketing values for the temperature and normalization profiles obtained in this fashion are shown in Fig.~\ref{fig:inst_profile}. Again, the trends suggest that the results are robust.

\section{Discussion}
\label{sect:discus}

We have presented the first well-constrained measurements of the ICM properties for a modestly sized, dynamically young galaxy cluster out to large radii. Here we compare our findings with the results obtained for hotter, more massive clusters, and with predictions based on numerical simulations. Finally, we discuss the implications. 

\subsection{The unrelaxed outskirts of a forming cluster}
\label{unrelaxed}

The surface brightness profile along the northern direction of the Virgo Cluster is unusually shallow. For 100~kpc~$<r<$~1200~kpc it follows a power-law shape $S_{\mathrm{X}}\propto r^{-\alpha}$ with index $\alpha=1.34\pm0.01$, which is significantly smaller than $\alpha\sim3$ for Abell~1795 \citep{2009PASJ...61.1117B} or $\alpha=3.6$ for a sample of massive luminous clusters observed with \chandra\ between 0.4--0.7 $r_{200}$, \citep{ettori2009}. Simulations of low mass clusters ($M_{\mathrm{vir}}<10^{15}$~M$_\odot$) by \citet{roncarelli2006} predict the ratio between the surface brightnesses at $r_{200}$ and $0.3r_{200}$ to be $\sim8\times10^{-3}$, which is more than an order of magnitude smaller than the value of 0.16 measured in Virgo. The emission in the outskirts of this young forming cluster is therefore considerably brighter than expected.

The derived density profile follows a power-law shape $n_{\mathrm{e}}\propto r^{-\beta}$ with index $\beta=1.21\pm0.12$. For comparison, the density profile of the Perseus cluster has $\beta=1.68\pm0.04$ \citep{simionescu2011} and that of Abell~1795 is even steeper ($\beta=2.27\pm0.07$). Simulations by \citet{roncarelli2006} predict that the gas density in the cluster outskirts (out to $\sim1.2\,r_{\text{vir}}$) can be described by a power-law model with an index $\beta=2.4-2.5$.

We measure a $\sim$60 per cent drop in temperature between $0.3-1r_{200}$, which is consistent with previous observations of some more massive relaxed clusters with \suzaku\ \citep{2009A&A...501..899R,2009PASJ...61.1117B,2010arXiv1001.5133H,simionescu2011}. In PKS~0745-191 the temperature drops by a factor of 4 \citep{2009MNRAS.395..657G}. 
Simulations \citep{roncarelli2006} predict a drop of $\sim$40 per cent, which is also consistent with the analytic result by \citet{ostriker2005}.

As we show in the middle panel of Fig.~\ref{fig:nesp} in Sect.~\ref{specres}, outside $r\sim450\,\text{kpc}$ the measured entropy is consistently lower, by a factor of 2--2.5, than values expected from hydrodynamic simulations of gravitationally collapsed gas in hydrostatic equilibrium, which predict $K\propto r^{1.1}$ \citep{tozzi2001,2005RvMP...77..207V}. Flattening entropy profiles have also been seen in PKS~0745-191 \citep{2009MNRAS.395..657G} and especially in the Perseus Cluster \citep{simionescu2011}; however, in these hotter systems the entropy profile starts to deviate from the expected shape at larger scaled radii. 

\citet{simionescu2011} show that the break in the measured entropy profile, and a simultaneous increase in the apparent gas mass fraction above the cosmic mean in the outskirts of the Perseus Cluster, are most likely caused by clumping in the ICM. The quantity that we infer from the observed deprojected emission measure is the average of the square of the electron density $\left<n^2\right>$, not the average of the electron density squared $\left<n\right>^2$. Therefore, if the gas is clumpy, its non-uniform density will lead to an overestimate of the average electron density calculated from the emission measure. If the dense clumps are in pressure equilibrium with the surrounding gas, then they will be cooler than the ambient medium. However, it is more likely that they will be falling into the cluster, moving through the ICM and therefore partly confined by ram-pressure.  Depending on the ratio of ram-pressure to thermal pressure support of the clumps, the average temperature in the multiphase regions will be underestimated because denser and cooler blobs have larger volume emissivity. Therefore, increasing clumping as a function of radius will result in a flattening of the observed surface brightness and density profiles and a steepening of the observed temperature profile. The over-estimated density and underestimated temperature at large radii will then combine into a flattening of the entropy profile, as observed. 

In order to test further whether clumping can be responsible for the observed features in our profiles, we simulated an \xmm\ spectrum of a two phase ICM, where both phases have a metallicity of $Z=0.3$~Solar. A hotter 2~keV component is used to represent the ambient ICM and a 1~keV phase represents the clumps of cooler plasma. The emission measures were adjusted so that, assuming thermal pressure equilibrium, the cooler clumps will have a volume filling fraction of $f=0.2$ (additional ram-pressure support will make the volume filling fraction of the clumps smaller). The simulated spectrum contains all the observed background components in the Virgo Cluster and has a similar statistical quality to our observed spectra from large radii. When fitting this spectrum with a single temperature thermal model we obtain a good fit, with a temperature of $kT=1.24$~keV and a metallicity of $Z=0.16$~Solar, underestimating the true metallicity by about a factor of 2. We also overestimate the density of the ICM by a factor of 1.6. The entropy inferred from this fit is about 2.2 times lower than the true entropy of the ICM, indicating that such multi-phase medium could indeed be responsible for the flattening entropy profile. The presence of cooler, denser blobs also explains the abrupt decrease of the temperature and observed metallicity beyond 450~kpc, where the entropy starts to deviate from the expected power-law shape. 

Clumping therefore appears to provide a natural explanation for the properties of the observed density, temperature, entropy, pressure, and metallicity profiles in the Virgo Cluster. We note that the real ICM is, however, probably multi-phase with different clumps having slightly different temperatures and the clumps being partly ram-pressure confined.

While X-ray observations provide the measurement of $\left<n^2\right>$, SZ observations depend linearly on $n$. Therefore, in regions with clumped ICM, we expect to see a discrepancy between the X-ray and the SZ signals.  
If there is no bias in measuring the temperature (e. g. the clumps are completely ram-pressure supported and have the same temperature as the ambient plasma) then this difference will provide an independent measurement of the clumping factor. Temperature biases arising from the multi-phase structure of the ICM may complicate the interpretation.  

\subsection{Metals at large radii}

We present a metallicity profile spanning from the cool core region all the way out to $r_{200}$. Our metallicity measurement is dominated by the line emission of the Fe-L complex and therefore primarily represents the Fe abundance.
Our data provide direct observational evidence that the ICM is enriched by metals all the way to $r_{200}$. In the range of $\sim$0.5--1$\:r_{\text{vir}}$, the observed metallicity profile is consistent with being  flat at a value of about $Z=0.1$ Solar.  Previously, the ICM metallicity at large radii has been obtained by \citet{fujita2008superwind} from a large region spanning 0.5--1$r_{200}$ in the compressed area between the clusters Abell 399/401 at the onset of a merger. They found that the ICM at large radii has been enriched to $Z\sim0.3$~Solar \citep[with respect to the Solar abundances of ][]{grevesse1998}. More recently, \citet{simionescu2011} showed that the outskirts of the Perseus Cluster have also been enriched to a similar metallicity of $Z\sim0.3$ Solar.

The gas within our analyzed region is, however, most probably multi-phase, consisting of multiple temperature components. It has been shown that the metallicity is often sensitive to the modeling of the underlying temperature structure and, if multi-temperature plasma is modeled with a single-temperature spectral model, the metallicity of gas can be significantly underestimated \citep[e.g.][see also our simulation in Sect.~\ref{unrelaxed}]{buote2000b}. This is especially true around the temperature of 1~keV, where the Fe-L complex is very sensitive to the underlying temperature structure. Our measured metallicity could thus be an underestimate of the true metallicity and should be considered a lower limit. 

Nevertheless, using the best fit metallicity values and assuming that both the gas density and the metal distribution are homogeneous, we calculate the cumulative Fe mass of the Virgo cluster. The total Fe mass obtained this way is $\sim4\times10^9~M_{\odot}$ within $0.1-1R_{200}$, approximately half of which resides outside of 0.5$R_{200}$. Because the metallicity is likely biased low and neither the gas density nor the distribution of metals are likely to be homogeneous, this metal mass is almost certainly an underestimate. We conclude that the total mass of Fe beyond 0.5$r{200}$ is at least $2\times10^9~M_{\odot}$.

The two dominant mechanisms by which metals pollute the ICM are galactic winds and ram-pressure stripping. Galactic winds, driven by the thermal energy of a large number of supernova explosions \citep[e.g.][]{heckman2003}, are expected to dominate the enrichment of the intergalactic medium at redshifts of $z=2$--3, when the bulk of the star formation happened. Part of this pre-enriched inter-galactic medium subsequently fell into the gravitational potential of clusters where it got shock heated and today constitutes the ICM. 

\citet{fujita2008superwind} interpret the high metallicity in the outskirts of Abell 399/401 as evidence for early enrichment by galactic super-winds. They argue that in order to exceed the pressure of $\sim2\times10^{-12}\,\text{Pa}$ necessary to start stripping the galaxies \citep{fujita1999ram} their velocities would have to exceed $\sim2000\,\text{km}\thinspace\text{s}^{-1}$. Following this argument, the relative velocities between the galaxies of the Virgo Cluster and the ICM would have to exceed the unrealistically high value of $\sim3000\,\text{km}\thinspace\text{s}^{-1}$. However, \ion{H}{i} imaging surveys have revealed several galaxies falling into the Virgo Cluster that display long \ion{H}{i} tails at projected radii of 0.6--1.2~Mpc from the center of the cluster \citep[e.g.][]{kenney2004,chung2007,kantharia2008}. There are furthermore several examples which suggest that stripping also operates in lower density, lower velocity dispersion environments of poor clusters and groups of galaxies \citep{kantharia2005,levy2007}. 
Recent simulations \citep[e.g.][]{roediger2007,roediger2008} show that galaxies can get stripped already at large radii with low ICM densities due to continuous or turbulent/viscous stripping \citep[see][]{nulsen1982b}. In principle, all the metals outside 0.5$R_{200}$ could be supplied by the complete stripping of  $\sim$150 galaxies, assuming $10^{10}~M_{\odot}$ of stripped gas, enriched to the Solar metallicity, per average galaxy.  
Stripping would, however, produce a radially decreasing metallicity profile at the large radii, while the observed profiles in both Virgo and Perseus are consistent with being flat.

The outskirts of Virgo have three times lower apparent metallicity than the outskirts of the Perseus Cluster and the compressed region between the clusters Abell 399/401. A similar ultra-low metallicity of $Z=0.15$~Solar has only been measured in the group of galaxies NGC~5044 at radii 0.2--0.4$r_{\mathrm{vir}}$ \citep{buote2004}. Given that the stellar over ICM baryon fraction of clusters is decreasing with the increasing cluster mass \citep[e.g.][]{gonzales2007} explaining the higher metal content of the more massive clusters would be challenging.
However, this difference may be primarily due to the multiphase gas biasing the Fe abundance low in cooler systems more strongly than in hotter systems. While the metallicity of Virgo at large radii has been determined based on the Fe-L lines, the metallicities of hotter clusters, $kT\gtrsim3.5$~keV, are determined based on their Fe-K lines. The metallicity of the hotter ICM could be biased both toward higher and lower values, depending on the temperature structure. The bias is however much smaller than in the cooler systems, only of the order of up to $\sim$30 per cent \citep[e.g.][]{rasia2008,simionescu2009b,gastaldello2010}.

\section{Conclusions}
\label{sect:concl}

We present the results of the analysis of a mosaic of \xmm\ observations of the modestly sized, dynamically young Virgo Cluster from the center to the north out to $\sim1.2\,\text{Mpc}$. We detect X-ray emission all the way out to this radius, which allows us to study the ICM properties in unprecedented detail. The surface brightness and density profiles are significantly shallower than predicted by simulations and measured in other more massive systems. 
The temperature is measured to drop by $\sim$60 per cent between $0.3-1\,r_{\text{vir}}$. Beyond the radius of $\sim$450~kpc the temperature and metallicity drop abruptly and the entropy profile deviates from the power-law shape $K\propto r^{1.1}$ expected for gravitationally collapsed gas in hydrostatic equilibrium \citep{tozzi2001,2005RvMP...77..207V}. At these radii, the entropy is consistently lower than expected by a factor of 2--2.5.

The most likely explanation for the unusually shallow density profile, the flattening of the entropy profile, and sharp drops in temperature and metallicity at $r>450$~kpc is the onset of significant gas clumping, which then increases as a function of radius. The clumping in the Virgo Cluster becomes significant at smaller scaled radii than has been detected in the more massive Perseus Cluster \citep{simionescu2011}.

Our data provide direct observational evidence that the ICM is significantly enriched with metals all the way to $r_{200}$. The measured metallicity profile flattens out in the cluster outskirts and reaches $Z=0.11\pm0.02$ Solar at $r_{200}$. The metallicity at large radii is, however, likely underestimated significantly because of the clumping and therefore this measured value should be considered a lower limit.     

The Virgo Cluster, being the closest galaxy cluster to us, provides a unique opportunity to study large-scale structure formation in detail as it happens. The similarities and differences between the results for the Virgo Cluster presented here and for the Perseus Cluster discussed by \citet{simionescu2011} point to a wealth of interesting physics at large radii of clusters that is just beginning to be probed. Understanding this physics presents an exciting challenge for numerical simulations.

\section*{Acknowledgements}
We thank Paul Nulsen and Marcus Br\"uggen for stimulating discussions. Support for this work was provided by the National Aeronautics and Space Administration through Chandra/Einstein Postdoctoral Fellowship Award Numbers PF8-90056 and PF9-00070 issued by the Chandra X-ray Observatory Center, which is operated by the Smithsonian Astrophysical Observatory for and on behalf of the National Aeronautics and Space Administration under contract NAS8-03060. This work was supported in part by the US Department of Energy under contract number DE-AC02-76SF00515. This work is based on observations obtained with \xmm, an ESA science mission with instruments
and contributions directly funded by ESA member states and the USA (NASA). 

\bibliographystyle{mnras}
\bibliography{clusters}

\end{document}